\documentclass[ aip,jap, jmp,amsmath,amssymb, reprint,]{revtex4-1}
\usepackage{graphicx}
\usepackage{dcolumn}
\usepackage{bm}
\usepackage{hyperref}

\begin{document}
\preprint{AIP/123-QED}
\title{Measurement of spin diffusion in semi-insulating GaAs}
\date{\today}
\author{C. P. Weber}\email{cweber@scu.edu}
\author{Craig A. Benko}
\altaffiliation[Present address: ]{JILA, National Institute of Standards and Technology, and University of Colorado, Department of Physics, 440 UCB, Boulder, CO 80309, USA}
\author{Stanley C. Hiew}
\affiliation{Department of Physics, Santa Clara University, 500 El Camino Real, Santa Clara, CA 95053-0315}

\begin{abstract}
We use optical transient-grating spectroscopy to measure spin diffusion of optically oriented electrons in bulk, semi-insulating GaAs(100). Trapping and recombination do not quickly deplete the photoexcited population. The spin diffusion coefficient of $88\pm12$ cm$^2$/s is roughly constant at temperatures from 15 K to 150 K, and the spin diffusion length is at least 450 nm. We show that it is possible to use spin diffusion to estimate the electron diffusion coefficient. Due to electron-electron interactions, the electron diffusion is 1.4 times larger than the spin diffusion.\\
\\
The following article appeared in \textit{Journal of Applied Physics} and may be found at \url{http://link.aip.org/link/?jap/109/106101} \\ \\
\textit{Copyright 2011 American Institute of Physics. This article may be downloaded for personal use only. 
Any other use requires prior permission of the author and the American Institute of Physics.}
\end{abstract}
\maketitle

The burgeoning field of semiconductor spintronics relies on moving spin-polarized electrons through distances comparable to the dimensions of an electronic device. The importance of spin transport has led to several studies of spin diffusion in GaAs quantum wells. Spin transport in quantum wells can differ markedly from the bulk material, due to to different scattering rates and especially to the different spin-orbit coupling \protect{\cite{Weber2007}}. Nonetheless, there have been relatively few measurements\protect{\cite{Kikkawa1999, Crooker2005, YuAPL2009}} of spin diffusion in bulk GaAs. In $n$-doped samples with $n=1\times 10^{16}$ and $2\times 10^{16}$ cm$^{-3}$, the spin diffusion coefficient $D_s$ ranged from 10 to 200 cm$^2$/s.

Spin diffusion in semi-insulating GaAs (SI-GaAs) has not been reported. SI-GaAs has been proposed as a platform for nuclear spintronics \protect{\cite{Reimer}} due to its low carrier density. Moreover, Kikkawa \textit{et al.} showed that electrons could be optically oriented in SI-GaAs and would subsequently diffuse into an adjacent ZnSe film, maintaining their spin polarization \protect{\cite{Kikkawa2001}}.  Since SI-GaAs is a ubiquitous substrate material for thin film growth and for spintronic devices, such spin diffusion is of practical consequence, whether intentional or not. In this work, we find that SI-GaAs has a large, temperature-independent spin diffusion coefficient.

We measured spin diffusion with an ultrafast transient spin grating \protect{\cite{Cameron1996}}, which measures the decay rate $\gamma_s$ of a spin-density wave (the ``grating'') with wavelength $\Lambda$ and wavevector $q=2\pi/\Lambda$. The grating amplitude decays---through spin relaxation, electron-hole recombination, and diffusion---at a rate of
\begin{equation}\gamma_s(q)=D_sq^2+1/\tau_0.
\label{DiffusionEq}
\end{equation}
Here, $D_s$ is the spin diffusion coefficient, and $\tau_0$ is the lifetime for trapping, recombination, and spin relaxation. Measurement at several $q$ determines $D_s$. We measure in a reflection geometry, and improve the detection efficiency by heterodyne detection \protect{\cite{Vohringer1995}}. Noise is further suppressed by 95 Hz modulation of the grating phase and lock-in detection \protect{\cite{Weber2005}}.

The SI-GaAs sample was grown by Wafer Technology. It was undoped, oriented (100), had room temperature resistivity $\rho\geq10^7$ $\Omega$-cm and Hall mobility $\mu_H\geq5000$ cm$^2/$V-s.

The pump and probe pulses came from a mode-locked Ti:Sapphire laser with wavelength near 800 nm and repetition rate of 80 MHz. The two pump pulses were focused to a spot of 65 $\mu$m diameter with total fluence 3.0 $\mu$J/cm$^2$ except as indicated; probe pulses were always a factor of 2.5 weaker. Assuming one photoexcited electron per absorbed photon in a 1 $\mu$m absorption length \protect{\cite{SturgePhysRev1962}}, we photoexcite $\sim8.5\times10^{16}$ cm$^{-3}$ carriers, greater than the typical concentration of deep traps in SI-GaAs \protect{\cite{MartinJAP1980}}. In this way we are able to measure motion of free carriers at times longer than the trapping time, and to use the density of photoexcited electrons as an estimate of the free-carrier density.

\begin{figure}
\protect{\includegraphics[width=2.5in]{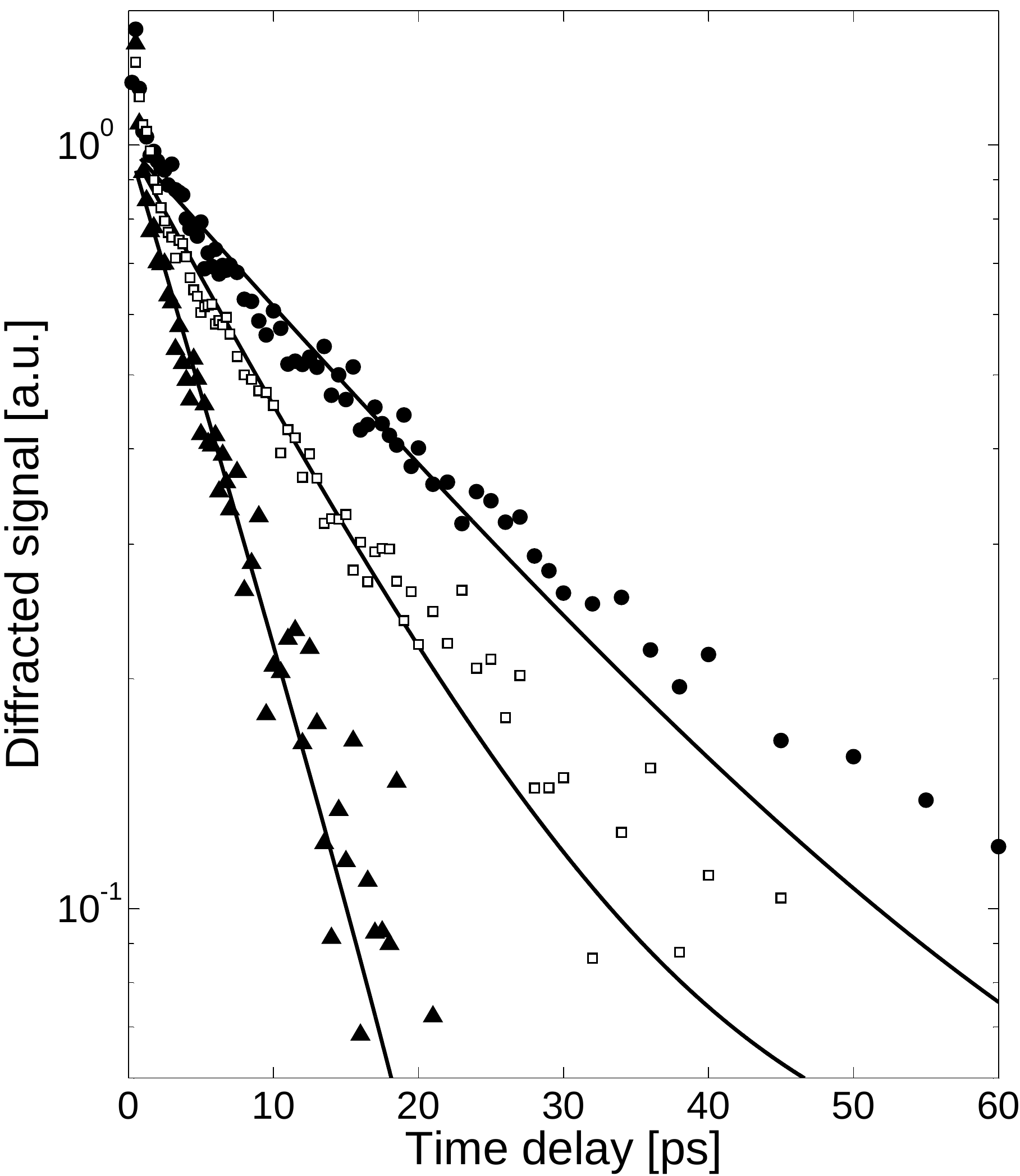}}
\caption{Decay of the transient spin grating at 15 K (semilog scale). Curves correspond to wavevectors $q=$ 2.01 $\times 10^4$, 3.14 $\times 10^4$, and 4.53 $\times 10^4$ cm$^{-1}$ (slowest to fastest). Solid lines are a least-squares fit to an exponential decay plus a constant offset.}
\label{Decays}
\end{figure}

Figure \ref{Decays} shows typical results of spin-grating measurements. After dropping rapidly for 0.5 ps, the diffracted signal---and the spin grating amplitude---decays exponentially at rate $\gamma_s$. Higher-$q$ gratings decay more quickly, as expected for diffusive motion. The solid lines show fits of the data to the form $A+B\exp[-\gamma_s(q)t]$. The size of the constant offset $A$ averages 2.5\% of the exponential decay (and never exceeds 7\%), so it does not significantly influence the values of $\gamma_s$. We speculate that the offset may arise from a small fraction of localized carriers.

\begin{figure}
\protect{\includegraphics[width=3in]{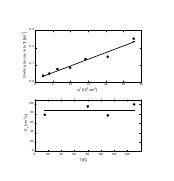}}
\caption{(a): Decay rate of the spin grating vs $q^2$ at 15 K. The line is a least-squares fit to the form of Eq. \ref{DiffusionEq}, indicating diffusive behavior with $D_s=78$ cm$^2$/s. (b): Spin diffusion coefficient vs temperature. The line is the mean value of 88 cm$^2$/s.}
\label{Diffusion}
\end{figure}

We determined spin diffusion coefficients by fitting $\gamma_s(q)$ to Eq. \ref{DiffusionEq}, as shown in Fig. \ref{Diffusion}a. The measured $D_s$ represents electron spin diffusion with no appreciable contribution from the holes. The near-bangap absorption of circularly polarized light by GaAs excites spin-polarized electrons and holes, but the hole spins rapidly randomize, leaving behind spin-aligned electrons \protect{\cite{OpticalOrientation}}. By fitting to times after the rapid initial decay, we obtain $\gamma_s$ for electrons only. Moreover, the photoexcited electron and hole populations are spatially uniform, so the electrons' motion is not hindered by electron-hole Coulomb attraction, and spin diffusion may occur more quickly than ambipolar diffusion \protect{\cite{Cameron1996}}. Uniform excitation also precludes any photorefractive grating.

The 15 K data in Figs. \ref{Decays} and \ref{Diffusion}a were all taken with pump fluence of 3.0 $\mu$J/cm$^2$. The signal size decreases at higher temperatures, so data were taken with fluences of up to 9.0 $\mu$J/cm$^2$. We found that fluence had a small effect on grating decay rate: as compared to 3.0 $\mu$J/cm$^2$, decay rates measured at 6.0 to 9.0 $\mu$J/cm$^2$ were typically slower by 5-20\%, while attenuating to 1.5 $\mu$J/cm$^2$ did not change $\gamma_s$ at all. The origin of this fluence-dependent grating decay is a topic of further research. However, the measured changes in $\gamma_s$ are sufficiently small that they do not add much uncertainty to the measured $D_s$.

Fig. \ref{Diffusion}b shows the values of $D_s$ at several temperatures as determined from the fits to Eq. \ref{DiffusionEq}. The spin diffusion is roughly constant in temperature, and it is fast---comparable to that previously seen\protect{\cite{Weber2005}} at low temperature in a quantum well with $\mu=69000$ cm$^2$/V-s. This supports our conclusion that most photoexcited electrons remain mobile for times of at least $\tau_0$. (The values of $\tau_0$ were 60, 100, 28, and 23 ps at 15, 80, 110, and 150 K, respectively.) Noting that the spin lifetime  $\tau_s\geq\tau_0$, the spin diffusion length $L_s=\sqrt{D_s\tau_s}$ is at least 450 nm. This length, comparable to the optical absorption length, suggests the possibility of efficient spin injection from SI-GaAs into thin epilayers \protect{\cite{Kikkawa2001}}.

Finally, we infer electron diffusion from the measured spin diffusion. One must consider spin Coulomb drag \protect{\cite{SCD}}, the effect of electron-electron collisions that transfer momentum between counter-diffusing spin-up and spin-down populations. This effect has been observed to suppress spin diffusion, relative to electron diffusion, by factors up to 8 in high-mobility quantum wells \protect{\cite{Weber2005}}. The electron diffusion is \protect{\cite{DAmicoPRB2002}}:
\begin{equation}
D_e=\frac{D_s}{\chi_0/\chi_s-\chi_0e^2\rho_{\uparrow\downarrow}D_s}.
\label{DefromDs}
\end{equation}
Here $\chi_s$ is the spin susceptibility, $\chi_0=\partial n/\partial\mu$ is the electronic susceptibility, and $\rho_{\uparrow\downarrow}$ is the spin transresistivity. Our inferred value of $D_e$ thus should be regarded as an estimate, since it depends---through $\chi_s$, $\chi_0$ and $\rho_{\uparrow\downarrow}$---on the density $n$ and heating $\Delta T$ of the photoexcited electrons, which are known only approximately\protect{\cite{HeatingNote}}.

We estimate $\rho_{\uparrow\downarrow}(T)$ from Fig. 1 of Ref. \protect{\onlinecite{DAmicoPRB2002}} and calculate $\chi_0(T)$ numerically \protect{\cite{Mohankumar1995}} for a noninteracting electron gas of density $n=8.5\times10^{16}$ cm$^{-3}$ assuming spherical, parabolic bands. We use the Perdew-Wang parametrization \protect{\cite{Perdew1992b}} for the exchange-correlation energy to obtain $\chi_0/\chi_s\approx0.82$. Eq. \ref{DefromDs} gives values of $D_e$ ranging from 110 to 140 cm$^2$/s. $D_e$ is consistently about 40\% higher than $D_s$, showing the importance of electron-electron interactions even in this high-resistivity material.

As a check on $D_e$, we convert diffusion to mobility using the Einstein relation, $\mu_e=e\chi_0D_e/n$. We find that $\mu_e$ increases from 7500 cm$^2$/V-s at the highest measured temperature to 13000 cm$^2$/V-s at the lowest. We are not aware of any measurement of electron mobility in SI-GaAs at low temperature. However, our mobility agrees reasonably with that of \textit{undoped} n-GaAs at low temperature \protect{\cite{Rode1975_2}}, while the room-temperature mobilities of SI-GaAs and of undoped n-GaAs are comparable \protect{\cite{Gooch1961, Rode1975_2, MartinJAP1980}}. This rough agreement shows that transient spin gratings can complement transport in measuring $D_e$. The technique might be particularly useful in ferromagnets with a large anomalous Hall effect.  

Our observation of roughly temperature-independent spin diffusion contrasts with the apparent strong temperature dependence in $n$-doped GaAs: a sample with  $n=2\times10^{16}$ cm$^{-3}$ had $D_s=10$ cm$^2$/s at 4 K (Ref.  \protect{\onlinecite{Crooker2005}}), while one with $n=1\times10^{16}$ cm$^{-3}$ had $D_s=200$ cm$^2$/s at room temperature \protect{\cite{YuAPL2009}}. In the absence of electron-electron interactions, the spin diffusion would equal the charge diffusion, $n\mu/e\chi_0$. Lowering temperature decreases $n/\chi_0$, but generally increases the mobility, two effects that partly cancel. At low $T$, $n/\chi_0$ approaches a nonzero minimum value due to degeneracy. Since our photoexcited density exceeds the densities studied in $n$-doped samples, degeneracy sets in at a higher temperature. Thus for the same $\mu(T)$ we would expect diffusion to in our samples to equal that of $n$-doped samples at 150 K, while exceeding it by a factor of 4 at low temperature. Differing mobility among our semi-insluating sample and the two $n$-doped samples reported likely accounts for the remaining differences in $D_s$.

Semi-insulating GaAs is an important substrate for spintronic materials and structures, and may also be useful for nuclear spintronics. We have measured its electron spin diffusion coefficient, $D_s$, under conditions in which most photoexcited carriers are not trapped. This coefficient, which is difficult to learn from transport measurements, measures the motion of electron spins. $D_s$ has a high value of $\sim88$ cm$^2$/s, independent of temperature. 

Our result suggests that SI-GaAs could be useful for injecting spin into adjacent layers. It also holds a lesson for optical-orientation experiments in which a film is measured on top of a SI-GaAs substrate: unless a barrier-layer is grown, spins in the substrate will likely diffuse into the material being measured. Finally, we have shown that one can use spin diffusion to estimate electron diffusion, if one takes proper account of electron degeneracy (through $\chi_0$), and of electron-electron interactions through the spin susceptibility and spin Coulomb drag. 

\begin{acknowledgements}
The authors thank G. Vignale for illuminating conversations and for sending a program to evaluate $\chi_0/\chi_s$. We thank Joe Orenstein for loaned equipment, R. A. Kaindl for a loaned Ti:Sapphire laser, and R. P. Campion for sending the sample. This research was supported by an award from Research Corporation.
\end{acknowledgements}


\begin{thebibliography}{19}%
\makeatletter
\providecommand \@ifxundefined [1]{%
 \@ifx{#1\undefined}
}%
\providecommand \@ifnum [1]{%
 \ifnum #1\expandafter \@firstoftwo
 \else \expandafter \@secondoftwo
 \fi
}%
\providecommand \@ifx [1]{%
 \ifx #1\expandafter \@firstoftwo
 \else \expandafter \@secondoftwo
 \fi
}%
\providecommand \natexlab [1]{#1}%
\providecommand \enquote  [1]{``#1''}%
\providecommand \bibnamefont  [1]{#1}%
\providecommand \bibfnamefont [1]{#1}%
\providecommand \citenamefont [1]{#1}%
\providecommand \href@noop [0]{\@secondoftwo}%
\providecommand \href [0]{\begingroup \@sanitize@url \@href}%
\providecommand \@href[1]{\@@startlink{#1}\@@href}%
\providecommand \@@href[1]{\endgroup#1\@@endlink}%
\providecommand \@sanitize@url [0]{\catcode `\\12\catcode `\$12\catcode
  `\&12\catcode `\#12\catcode `\^12\catcode `\_12\catcode `\%12\relax}%
\providecommand \@@startlink[1]{}%
\providecommand \@@endlink[0]{}%
\providecommand \url  [0]{\begingroup\@sanitize@url \@url }%
\providecommand \@url [1]{\endgroup\@href {#1}{\urlprefix }}%
\providecommand \urlprefix  [0]{URL }%
\providecommand \Eprint [0]{\href }%
\providecommand \doibase [0]{http://dx.doi.org/}%
\providecommand \selectlanguage [0]{\@gobble}%
\providecommand \bibinfo  [0]{\@secondoftwo}%
\providecommand \bibfield  [0]{\@secondoftwo}%
\providecommand \translation [1]{[#1]}%
\providecommand \BibitemOpen [0]{}%
\providecommand \bibitemStop [0]{}%
\providecommand \bibitemNoStop [0]{.\EOS\space}%
\providecommand \EOS [0]{\spacefactor3000\relax}%
\providecommand \BibitemShut  [1]{\csname bibitem#1\endcsname}%
\let\auto@bib@innerbib\@empty
\bibitem [{\citenamefont {Weber}\ \emph {et~al.}(2007)\citenamefont {Weber},
  \citenamefont {Orenstein}, \citenamefont {Bernevig}, \citenamefont {Zhang},
  \citenamefont {Stephens},\ and\ \citenamefont {Awschalom}}]{Weber2007}%
  \BibitemOpen
  \bibfield  {author} {\bibinfo {author} {\bibfnamefont {C.~P.}\ \bibnamefont
  {Weber}}, \bibinfo {author} {\bibfnamefont {J.}~\bibnamefont {Orenstein}},
  \bibinfo {author} {\bibfnamefont {B.~A.}\ \bibnamefont {Bernevig}}, \bibinfo
  {author} {\bibfnamefont {S.~C.}\ \bibnamefont {Zhang}}, \bibinfo {author}
  {\bibfnamefont {J.}~\bibnamefont {Stephens}}, \ and\ \bibinfo {author}
  {\bibfnamefont {D.~D.}\ \bibnamefont {Awschalom}},\ }\href@noop {} {\bibfield
   {journal} {\bibinfo  {journal} {Physical Review Letters}\ }\textbf {\bibinfo
  {volume} {98}},\ \bibinfo {pages} {4} (\bibinfo {year} {2007})}
  
\bibitem [{\citenamefont {Kikkawa}\ and\ \citenamefont
  {Awschalom}(1999)}]{Kikkawa1999}%
  \BibitemOpen
  \bibfield  {author} {\bibinfo {author} {\bibfnamefont {J.~M.}\ \bibnamefont
  {Kikkawa}}\ and\ \bibinfo {author} {\bibfnamefont {D.~D.}\ \bibnamefont
  {Awschalom}},\ }\href@noop {} {\bibfield  {journal} {\bibinfo  {journal}
  {Nature}\ }\textbf {\bibinfo {volume} {397}},\ \bibinfo {pages} {139}
  (\bibinfo {year} {1999})}
  
\bibitem [{\citenamefont {Crooker}\ \emph {et~al.}(2005)\citenamefont
  {Crooker}, \citenamefont {Furis}, \citenamefont {Lou}, \citenamefont
  {Adelmann}, \citenamefont {Smith}, \citenamefont {Palmstrom},\ and\
  \citenamefont {Crowell}}]{Crooker2005}%
  \BibitemOpen
  \bibfield  {author} {\bibinfo {author} {\bibfnamefont {S.~A.}\ \bibnamefont
  {Crooker}}, \bibinfo {author} {\bibfnamefont {M.}~\bibnamefont {Furis}},
  \bibinfo {author} {\bibfnamefont {X.}~\bibnamefont {Lou}}, \bibinfo {author}
  {\bibfnamefont {C.}~\bibnamefont {Adelmann}}, \bibinfo {author}
  {\bibfnamefont {D.~L.}\ \bibnamefont {Smith}}, \bibinfo {author}
  {\bibfnamefont {C.~J.}\ \bibnamefont {Palmstrom}}, \ and\ \bibinfo {author}
  {\bibfnamefont {P.~A.}\ \bibnamefont {Crowell}},\ }\href@noop {} {\bibfield
  {journal} {\bibinfo  {journal} {Science}\ }\textbf {\bibinfo {volume}
  {309}},\ \bibinfo {pages} {2191} (\bibinfo {year} {2005})}
  
\bibitem [{\citenamefont {Yu}\ \emph {et~al.}(2009)\citenamefont {Yu},
  \citenamefont {Zhang}, \citenamefont {Wang}, \citenamefont {Ni},
  \citenamefont {Niu},\ and\ \citenamefont {Lai}}]{YuAPL2009}%
  \BibitemOpen
  \bibfield  {author} {\bibinfo {author} {\bibfnamefont {H.-L.}\ \bibnamefont
  {Yu}}, \bibinfo {author} {\bibfnamefont {X.-M.}\ \bibnamefont {Zhang}},
  \bibinfo {author} {\bibfnamefont {P.-F.}\ \bibnamefont {Wang}}, \bibinfo
  {author} {\bibfnamefont {H.-Q.}\ \bibnamefont {Ni}}, \bibinfo {author}
  {\bibfnamefont {Z.-C.}\ \bibnamefont {Niu}}, \ and\ \bibinfo {author}
  {\bibfnamefont {T.}~\bibnamefont {Lai}},\ }\href@noop {} {\bibfield
  {journal} {\bibinfo  {journal} {Applied Physics Letters}\ }\textbf {\bibinfo
  {volume} {94}},\ \bibinfo {pages} {202109} (\bibinfo {year}
  {2009})}
  
\bibitem [{\citenamefont {Reimer}(2010)}]{Reimer}%
  \BibitemOpen
  \bibfield  {author} {\bibinfo {author} {\bibfnamefont {J.~A.}\ \bibnamefont
  {Reimer}},\ }\href@noop {} {\bibfield  {journal} {\bibinfo  {journal} {Solid
  State Nuclear Magnetic Resonance}\ }\textbf {\bibinfo {volume} {37}},\
  \bibinfo {pages} {3} (\bibinfo {year} {2010})}
  
\bibitem [{\citenamefont {Kikkawa}\ \emph {et~al.}(2001)\citenamefont
  {Kikkawa}, \citenamefont {Gupta}, \citenamefont {Malajovich},\ and\
  \citenamefont {Awschalom}}]{Kikkawa2001}%
  \BibitemOpen
  \bibfield  {author} {\bibinfo {author} {\bibfnamefont {J.~M.}\ \bibnamefont
  {Kikkawa}}, \bibinfo {author} {\bibfnamefont {J.~A.}\ \bibnamefont {Gupta}},
  \bibinfo {author} {\bibfnamefont {I.}~\bibnamefont {Malajovich}}, \ and\
  \bibinfo {author} {\bibfnamefont {D.~D.}\ \bibnamefont {Awschalom}},\
  }\href@noop {} {\bibfield  {journal} {\bibinfo  {journal} {Physica E:
  Low-Dimensional Systems \& Nanostructures}\ }\textbf {\bibinfo {volume}
  {9}},\ \bibinfo {pages} {194} (\bibinfo {year} {2001})}
  
\bibitem [{\citenamefont {Cameron}, \citenamefont {Riblet},\ and\ \citenamefont
  {Miller}(1996)}]{Cameron1996}%
  \BibitemOpen
  \bibfield  {author} {\bibinfo {author} {\bibfnamefont {A.~R.}\ \bibnamefont
  {Cameron}}, \bibinfo {author} {\bibfnamefont {P.}~\bibnamefont {Riblet}}, \
  and\ \bibinfo {author} {\bibfnamefont {A.}~\bibnamefont {Miller}},\
  }\href@noop {} {\bibfield  {journal} {\bibinfo  {journal} {Physical Review
  Letters}\ }\textbf {\bibinfo {volume} {76}},\ \bibinfo {pages} {4793}
  (\bibinfo {year} {1996})}
  
\bibitem [{\citenamefont {Vohringer}\ and\ \citenamefont
  {Scherer}(1995)}]{Vohringer1995}%
  \BibitemOpen
  \bibfield  {author} {\bibinfo {author} {\bibfnamefont {P.}~\bibnamefont
  {Vohringer}}\ and\ \bibinfo {author} {\bibfnamefont {N.~F.}\ \bibnamefont
  {Scherer}},\ }\href@noop {} {\bibfield  {journal} {\bibinfo  {journal}
  {Journal of Physical Chemistry}\ }\textbf {\bibinfo {volume} {99}},\ \bibinfo
  {pages} {2684} (\bibinfo {year} {1995})}
  
\bibitem [{\citenamefont {Weber}\ \emph {et~al.}(2005)\citenamefont {Weber},
  \citenamefont {Gedik}, \citenamefont {Moore}, \citenamefont {Orenstein},
  \citenamefont {Stephens},\ and\ \citenamefont {Awschalom}}]{Weber2005}%
  \BibitemOpen
  \bibfield  {author} {\bibinfo {author} {\bibfnamefont {C.~P.}\ \bibnamefont
  {Weber}}, \bibinfo {author} {\bibfnamefont {N.}~\bibnamefont {Gedik}},
  \bibinfo {author} {\bibfnamefont {J.~E.}\ \bibnamefont {Moore}}, \bibinfo
  {author} {\bibfnamefont {J.}~\bibnamefont {Orenstein}}, \bibinfo {author}
  {\bibfnamefont {J.}~\bibnamefont {Stephens}}, \ and\ \bibinfo {author}
  {\bibfnamefont {D.~D.}\ \bibnamefont {Awschalom}},\ }\href@noop {} {\bibfield
   {journal} {\bibinfo  {journal} {Nature}\ }\textbf {\bibinfo {volume}
  {437}},\ \bibinfo {pages} {1330} (\bibinfo {year} {2005})}
  
\bibitem [{\citenamefont {Sturge}(1962)}]{SturgePhysRev1962}%
  \BibitemOpen
  \bibfield  {author} {\bibinfo {author} {\bibfnamefont {M.~D.}\ \bibnamefont
  {Sturge}},\ }\href@noop {} {\bibfield  {journal} {\bibinfo  {journal}
  {Physical Review}\ }\textbf {\bibinfo {volume} {127}},\ \bibinfo {pages}
  {768} (\bibinfo {year} {1962})}
  
\bibitem [{\citenamefont {Martin}\ \emph {et~al.}(1980)\citenamefont {Martin},
  \citenamefont {Farges}, \citenamefont {Jacob}, \citenamefont {Hallais},\ and\
  \citenamefont {Poiblaud}}]{MartinJAP1980}%
  \BibitemOpen
  \bibfield  {author} {\bibinfo {author} {\bibfnamefont {G.~M.}\ \bibnamefont
  {Martin}}, \bibinfo {author} {\bibfnamefont {J.~P.}\ \bibnamefont {Farges}},
  \bibinfo {author} {\bibfnamefont {G.}~\bibnamefont {Jacob}}, \bibinfo
  {author} {\bibfnamefont {J.~P.}\ \bibnamefont {Hallais}}, \ and\ \bibinfo
  {author} {\bibfnamefont {G.}~\bibnamefont {Poiblaud}},\ }\href@noop {}
  {\bibfield  {journal} {\bibinfo  {journal} {Journal of Applied Physics}\
  }\textbf {\bibinfo {volume} {51}},\ \bibinfo {pages} {2840} (\bibinfo {year}
  {1980})}
  
\bibitem [{\citenamefont {Meier}\ and\ \citenamefont
  {Zakharchenya}(1984)}]{OpticalOrientation}%
  \BibitemOpen
  \bibfield  {author} {\bibinfo {author} {\bibfnamefont {F.}~\bibnamefont
  {Meier}}\ and\ \bibinfo {author} {\bibfnamefont {B.}~\bibnamefont
  {Zakharchenya}},\ }\href@noop {} {\emph {\bibinfo {title} {Optical
  Orientation}}}\ (\bibinfo  {publisher} {North-Holland},\ \bibinfo {address}
  {Amsterdam},\ \bibinfo {year} {1984})
  
\bibitem [{\citenamefont {D'Amico}\ and\ \citenamefont {Vignale}(2001)}]{SCD}%
  \BibitemOpen
  \bibfield  {author} {\bibinfo {author} {\bibfnamefont {I.}~\bibnamefont
  {D'Amico}}\ and\ \bibinfo {author} {\bibfnamefont {G.}~\bibnamefont
  {Vignale}},\ }\href@noop {} {\bibfield  {journal} {\bibinfo  {journal}
  {Europhysics Letters}\ }\textbf {\bibinfo {volume} {55}},\ \bibinfo {pages}
  {566} (\bibinfo {year} {2001})}
  
\bibitem [{\citenamefont {D'Amico}\ and\ \citenamefont
  {Vignale}(2002)}]{DAmicoPRB2002}%
  \BibitemOpen
  \bibfield  {author} {\bibinfo {author} {\bibfnamefont {I.}~\bibnamefont
  {D'Amico}}\ and\ \bibinfo {author} {\bibfnamefont {G.}~\bibnamefont
  {Vignale}},\ }\href@noop {} {\bibfield  {journal} {\bibinfo  {journal}
  {Physical Review B-Condensed Matter}\ }\textbf {\bibinfo {volume} {65}},\
  \bibinfo {pages} {085109/1} (\bibinfo {year} {2002})}
  
\bibitem [{Hea()}]{HeatingNote}%
  \BibitemOpen
  \href@noop {} {}\bibinfo {note} {In what follows we assume electronic heating
  of $\Delta T=30$ K. We have checked that the value of $D_e$ depends only
  weakly (8\%) on $\Delta T$ in the range 0 K $\leq\Delta T\leq200$ K. Varying
  $n$ from $8.5\times10^{15}$ to $1.2\times10^{17}$ cm$^{-3}$ changes $D_e$ by
  no more than 7\%. The inferred $\mu_e$, on the other hand, varies strongly
  with $n$ and roughly as $1/(T+\Delta T)$.}
  
\bibitem [{\citenamefont {Mohankumar}\ and\ \citenamefont
  {Natarajan}(1995)}]{Mohankumar1995}%
  \BibitemOpen
  \bibfield  {author} {\bibinfo {author} {\bibfnamefont {N.}~\bibnamefont
  {Mohankumar}}\ and\ \bibinfo {author} {\bibfnamefont {A.}~\bibnamefont
  {Natarajan}},\ }\href@noop {} {\bibfield  {journal} {\bibinfo  {journal}
  {Physica Status Solidi B-Basic Research}\ }\textbf {\bibinfo {volume}
  {188}},\ \bibinfo {pages} {635} (\bibinfo {year} {1995})}
  
\bibitem [{\citenamefont {Perdew}\ and\ \citenamefont
  {Wang}(1992)}]{Perdew1992b}%
  \BibitemOpen
  \bibfield  {author} {\bibinfo {author} {\bibfnamefont {J.~P.}\ \bibnamefont
  {Perdew}}\ and\ \bibinfo {author} {\bibfnamefont {Y.}~\bibnamefont {Wang}},\
  }\href@noop {} {\bibfield  {journal} {\bibinfo  {journal} {Physical Review
  B}\ }\textbf {\bibinfo {volume} {45}},\ \bibinfo {pages} {13244} (\bibinfo
  {year} {1992})}
  
\bibitem [{Rod()}]{Rode1975_2}%
  \BibitemOpen
  \href@noop {} {}\bibinfo {note} {Rode, D. L., in \textit{Semiconductors and
  Semimetals}, Vol. 10: Transport Phenomena, edited by R. K. Willardson and A.
  C. Beer (Academic Press, 1975).}
  
\bibitem [{\citenamefont {Gooch}, \citenamefont {Hilsum},\ and\ \citenamefont
  {Holeman}(1961)}]{Gooch1961}%
  \BibitemOpen
  \bibfield  {author} {\bibinfo {author} {\bibfnamefont {C.~H.}\ \bibnamefont
  {Gooch}}, \bibinfo {author} {\bibfnamefont {C.}~\bibnamefont {Hilsum}}, \
  and\ \bibinfo {author} {\bibfnamefont {B.~R.}\ \bibnamefont {Holeman}},\
  }\href@noop {} {\bibfield  {journal} {\bibinfo  {journal} {Journal of Applied
  Physics}\ }\textbf {\bibinfo {volume} {32}},\ \bibinfo {pages} {2069}
  (\bibinfo {year} {1961})}
  
\end{thebibliography}
\end{document}